\documentclass[prl,aps,twocolumn,longbibliography]{revtex4-1}
\usepackage{color}
\usepackage{graphicx}
\usepackage{bm}
\usepackage{amsmath}
\usepackage{amssymb}
\usepackage{amsthm}
\usepackage{amsfonts}
\usepackage{enumerate}
\usepackage{latexsym}

\usepackage[colorlinks=true,urlcolor=blue,citecolor=blue,allcolors=blue]{hyperref}
\newcommand{\bs}{\boldsymbol}

\usepackage{braket}
\usepackage{array}
\usepackage{float}
\usepackage{cleveref}

\begin{document}

\title{Topological Random Fractals}
\author{Moein N.~Ivaki, Isac Sahlberg, Kim P\"oyh\"onen and  Teemu Ojanen}
\affiliation{Computational Physics Laboratory, Physics Unit, Faculty of Engineering and
Natural Sciences, Tampere University, P.O. Box 692, FI-33014 Tampere, Finland}
\affiliation{Helsinki Institute of Physics P.O. Box 64, FI-00014, Finland}

\begin{abstract}
We introduce the notion of topological electronic states on random lattices in non-integer dimensions. By considering a class $D$ model on critical percolation clusters embedded in two dimensions, we demonstrate that these topological random fractals exhibit a robust mobility gap, support quantized conductance and represent a well-defined thermodynamic phase of matter. The finite-size scaling analysis further suggests that the critical properties are not consistent with the class $D$ systems in two dimensions. Our results establish topological random fractals as the most complex systems known to support nontrivial band topology with their distinct unique properties.
\end{abstract}

\maketitle

\emph{Introduction---}
 Since the discovery of the quantum Hall effect, the quantized conductance, dissipationless currents and unconventional edge excitations have captured the fascination of generations of physicists~\cite{PhysRevLett.45.494,PhysRevLett.49.405,physRevB.23.5632,PhysRevLett.61.2015,PhysRevB.27.7539,PhysRevB.29.3303,PhysRevLett.71.3697,huckestein1995scaling}. These remarkable properties, unlikely from the point of view of traditional solid state physics, ultimately result from the topology of the electronic spectrum~\cite{RevModPhys.82.3045,RevModPhys.83.1057,ando2013topological,bernevig2013topological}. Recent efforts have revealed that topological states in naturally occurring materials are ubiquitous in nature~\cite{vergniory2019complete,zhang2019catalogue,tang2019comprehensive}.

Currently, the research of topological states of matter has moved beyond crystalline solids to amorphous and quasicrystalline systems~\cite{zhou2020amorphous,corbae2019evidence,mitchell2018amorphous,PhysRevLett.116.257002,Agarwala2017prl,yang2019metal,costa2019toward,PhysRevLett.121.126401,PhysRevLett.123.196401,mukati2020topological,agarwala2020higher,PhysRevE.104.025007,PhysRevLett.109.106402,araujo2019conventional,PhysRevLett.124.036803,PhysRevLett.111.226401,mitchell2018amorphous,poyhonen2018amorphous,PhysRevB.96.100202,mano2019application,sahlberg2020topological,Marsal30260,grushin2020topological,focassio2021amorphous,PhysRevX.6.011016,PhysRevResearch.2.043301,PhysRevX.9.021054,loring2019bulk,PhysRevB.103.214203,PhysRevLett.110.076403,PhysRevB.103.214203,hua2020higher,li2021symmetry,PhysRevLett.126.206404,10.21468/SciPostPhys.11.2.022,else2021quantum}. While it is not yet clear whether these states of matter exist in nature, they can be realized in artificial designer systems~\cite{lv2021realization,khajetoorians2019creating,drost2017topological,kempkes2019design,PhysRevResearch.3.023056}. Besides offering new avenues for functional devices, these systems open a new chapter in the physics of topological matter and the theory of Anderson localization. In this vein, the possibility of topological states in fractals has stirred a new research direction. Despite reported signatures of topology in a number of fractal lattices, many aspect of these systems remains unclear or controversial~\cite{fischer2021robustness,PhysRevResearch.2.013044,PhysRevB.104.045147,PhysRevB.100.155135,PhysRevB.101.045413,PhysRevB.98.205116,agarwala2018fractalized,song2014topological,manna2021laughlin,manna2021higher}. The existence and nature of the spectral gap, possibility of supporting quantized responses and anomalous dependence on system details (such as the coordination number of lattice sites) remain under debate. Furthermore, since the studies are mostly restricted to modest-size structures without systematic finite-size scaling analysis, it is not clear whether the finite samples actually represent a well-defined thermodynamic phase of matter. 
\begin{figure}
\includegraphics[width=0.65\columnwidth]{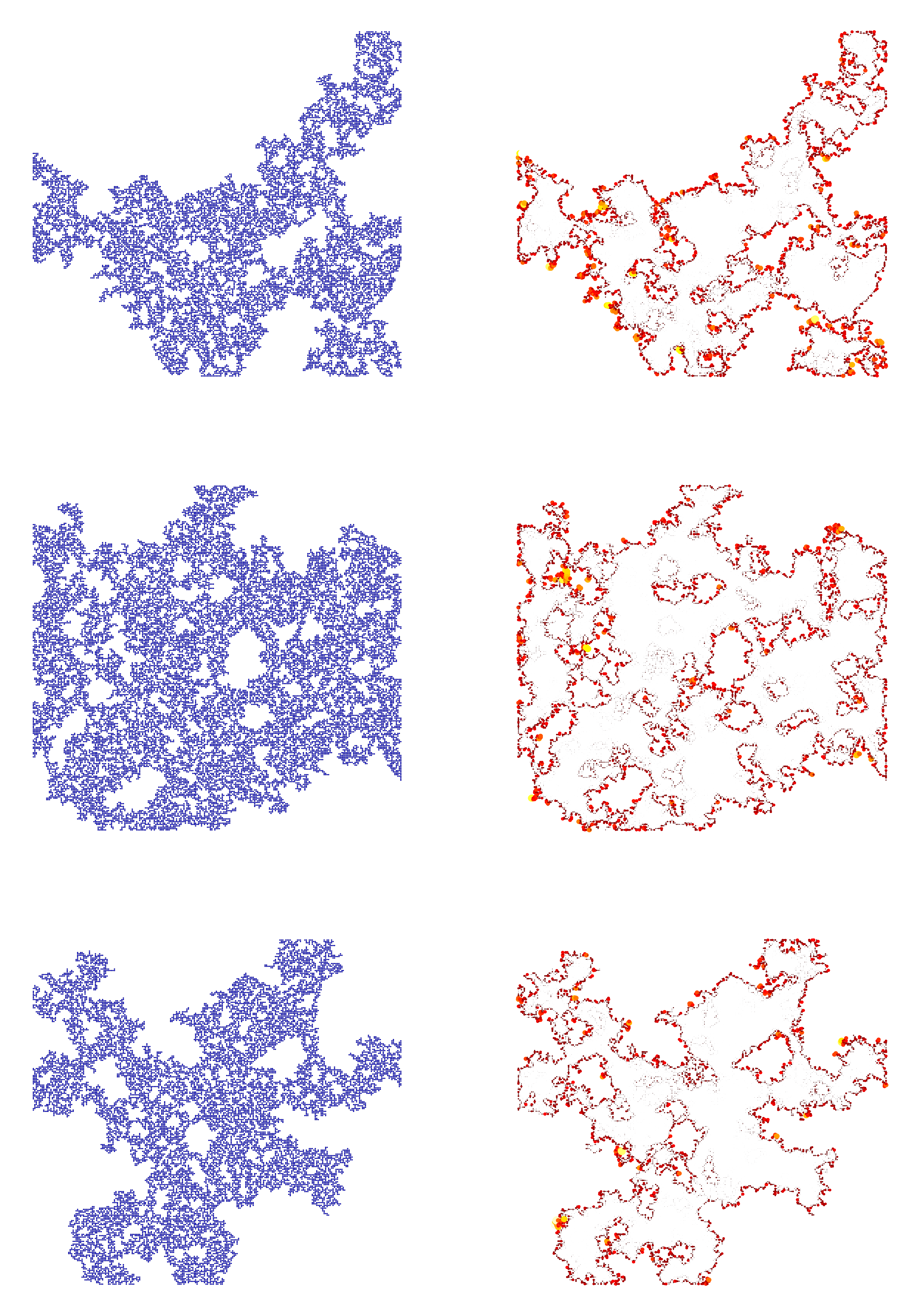}
\caption{Three finite snapshots with linear size $L=250$ of the studied random fractal lattices (left) and the local density of states of the topological model at the center of the mobility gap $E=0$ (right). Despite the seeming absence of clear distinction between the bulk and edge modes, these structures support robust quantized conductance.}
\label{fig:fractals_fig1}
\end{figure}

\begin{figure}
\includegraphics[width=0.8\columnwidth]{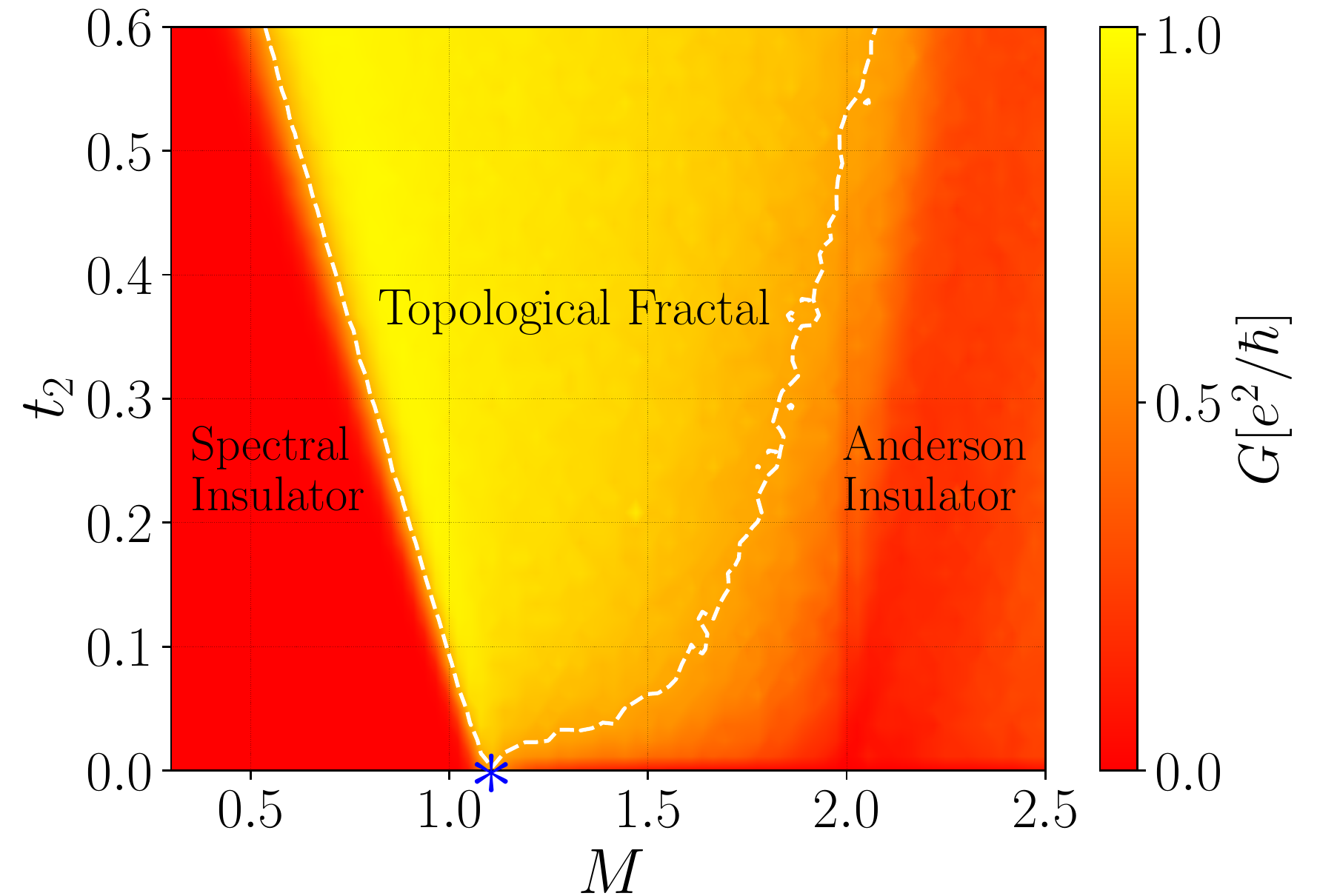}
\caption{Phase diagram of the studied system as a function of the second-nearest-neighbour hopping $t_2$ and the mass parameter $M$, obtained by calculation of the conductance $G$ for half-filled square-shaped systems with linear dimensions $L=150$. The dashed white line represents the approximate critical surface $G^{c}=0.65$, which encloses the topological regime where $G\to 1$ in the thermodynamic limit $L\to\infty$. The blue star signifies the tricritical point at $t_2=0$ and $M^c\approx1.1$. Data is averaged over 650 independent random realizations. }
\label{fig:fractal_pd}
\end{figure}

The research on topological fractals so far has been limited to deterministic self-similar structures. The organization principle of these structures, as of quasicrystals, is completely deterministic without any element of randomness. In contrast, in this work we demonstrate a novel topological phase on fundamentally more complex self-similar random lattices depicted in Fig.~\ref{fig:fractals_fig1}. These random fractals are \emph{statistically} self-similar, i.e. generated from a probability distribution, and characterized by a non-integer spatial dimension~\cite{nakayama2003fractal,PhysRevB.95.104206}. Specifically, we study a symmetry class $D$ Hamiltonian on critical clusters of 2d square lattice percolation. The geometry of the critical cluster is characterized by the fractal dimension $d_f=\frac{91}{48}\,\textless\,2$ (the number of sites within a circle of radius $r$ scales as $r^{d_f} $ for large $r$) and a set of standard critical exponents.
Our main findings are summarized in the following discoveries: I) the studied topological random fractals have in general a gapless energy spectrum but exhibit a well-defined mobility gap protecting the topological phase, II) the studied system supports robust quantized conductance, III) finite-size scaling analysis show that topological random fractals represent a well-defined thermodynamic phase of matter, IV) the localization exponent for class D random fractals is incompatible with the universal value $\nu=1$ in two dimensions. The last property suggests that, despite similarities with topological insulators in integer dimensions where they are embedded, topological random fractals represent a distinct state of matter.

\emph{Model and phase diagram---} In this work we consider random lattices arising from a site percolation process on a square lattice where each site is randomly occupied by probability $p$~\cite{stauffer2018introduction}. There exists a critical concentration $0\!<\!p_c\!<\!1$, known as the percolation threshold, above which the random lattice has an infinite nearest-neighbour-connected cluster in the thermodynamic limit. Below the threshold, the system consists of disconnected finite clusters. When approaching the percolation threshold, the characteristic length scale of the lattice $\xi$ (the geometric correlation length) diverges as $\xi \propto |p-p_c|^{-4/3}$, signifying that the percolating critical cluster becomes a scale-free  fractal. On the square lattice this takes place at $p_c\approx0.593$. The percolation transition at $p_c$ is sharply defined for an infinite system and the correlation length is understood as a statistical average. In Fig.~\ref{fig:fractals_fig1} we have illustrated three finite-size realizations of the random fractals generated by the above-described percolation process.

Next we define a two-band tight-binding model on critical square lattice percolation clusters. The model is determined by the Hamiltonian  
\begin{equation}\label{Hamiltonian}
        \begin{aligned}
            \mathbb{H}=\, &(2-M)\sum_{i}\bs{c}^{\dagger}_{i}\,\sigma_z\,\bs{c}_{i}\\
            &-\frac{t}{2} \sum_{<ij>}\bs{c}^{\dagger}_{i}\,\eta_{ij}\,\bs{c}_{j}
            -\frac{t_2}{2} \sum_{\ll ij\gg}\bs{c}^{\dagger}_{i}\,\eta_{ij}\,\bs{c}_{j} + \mathrm{h.c.},
        \end{aligned}
    \end{equation}
where $M$ is the onsite mass parameter and $t$ and $t_2$ represent the hopping amplitudes between the nearest- and second-nearest-neighbor sites on a square lattice provided those sites are present in a given random realization. The matrix $\eta_{ij}=\sigma_z + i\cos{\theta_{ij}}\sigma_x +i\sin{\theta_{ij}}\sigma_y$ is determined by $\theta_{ij}$, which denotes the angle between the $x$ axis and the bond vector from site $i$ to site $j$.  The two-component operators $\bs{c}^{\dagger}_{i}=(c_{i,1}^{\dagger}, c_{i,2}^{\dagger})$ create fermions at site $i$, and ${\sigma_{x,y,z}}$ are the Pauli matrices operating in the two-orbital space. The model \eqref{Hamiltonian} breaks time-reversal symmetry and satisfies particle-hole symmetry as $\sigma_x \mathbb{H}^* \sigma_x=-\mathbb{H}$, hence belonging to the symmetry class $D$ \cite{PhysRevB.55.1142,RevModPhys.88.035005}. On a square lattice with only nearest-neighbor hopping ($t_2=0$), the model supports a topological phase in the region $0<M/t<2 \,(2<M/t<4)$, classified by the Chern number $1\,(-1)$. In the remainder of this paper, we set $t=1$ and express the other parameters in units of $t$. On a regular lattice, a disordered class $D$ model is known to host a metallic phase which separates the two insulating topological phases. In the case of anisotropic models, an intervening localized phase appears. The localization exponents at metal-insulator and insulator-insulator transitions are known to be $\nu_{MI}\approx 1.4$ and $\nu_{II}= 1$, respectively~\cite{RevModPhys.80.1355,PhysRevB.81.214203,PhysRevLett.105.046803,PhysRevLett.101.127001,PhysRevB.65.012506,wang2021multicriticality,pan2021renormalization}.

\begin{figure}
\includegraphics[width=1.\columnwidth]{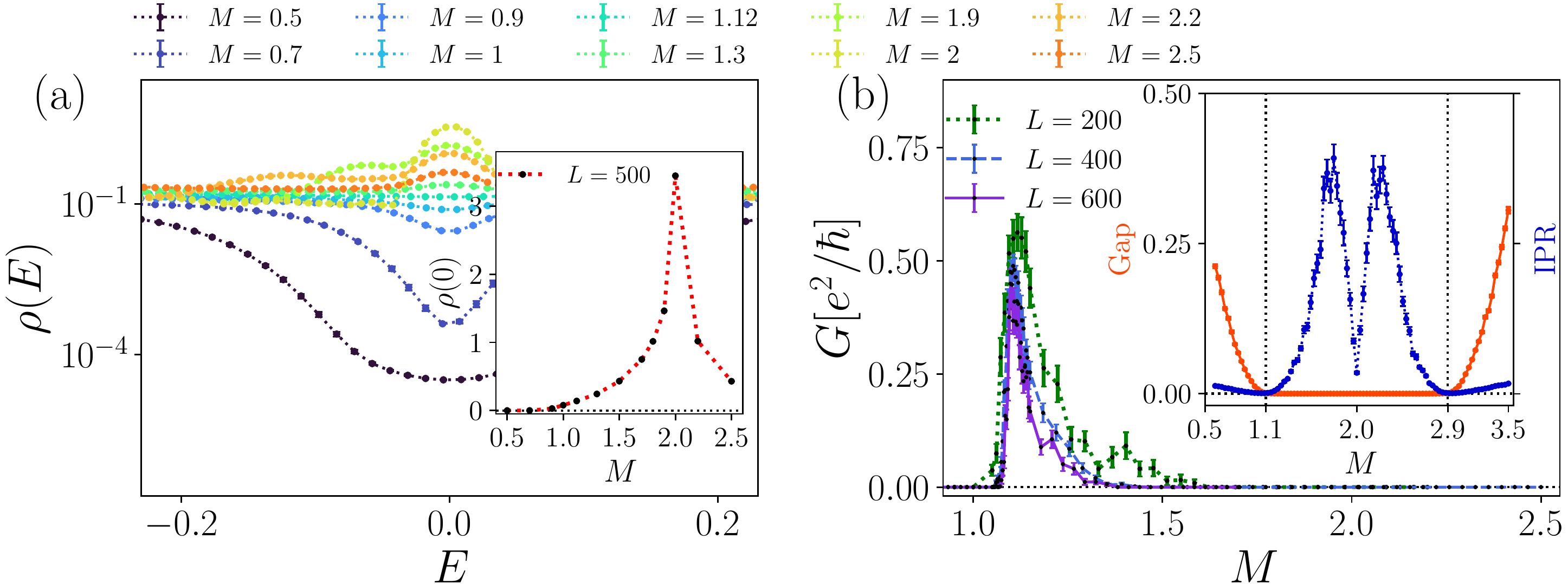}
\caption{Spectral insulator-Anderson insulator transition at $t_2=0$. (a) Configuration-averaged normalized DOS $\rho(E)$ around zero energy for the Hamiltonian~\eqref{Hamiltonian} for different $M$ values at $L=500$ and $t_2=0$. Inset shows the behavior of $\rho(E=0)$ as $M$ is varied. The feature at $M=2$ originates from a gapless point of the Hamiltonian~\eqref{Hamiltonian} on a square lattice. (b) Configuration-averaged Hall conductance, and the averaged spectral gap and IPR (inset) for $L=90$ as a function of $M$ at $t_2=0$. $M^c\approx1.1$ and $M^c\approx2.9$ denote the gapped-gapless transition points, with the gapless region hosting localized states.}
\label{fig:fig3}
\end{figure}
To numerically evaluate a two-terminal conductance $G$ we employ the KWANT software~\cite{groth2014kwant}, which implements transport calculations using scattering theory~\cite{RevModPhys.69.731}. It is assumed that the studied $L\times L$ samples are attached to two identical semi-infinite metallic leads, represented by decoupled 1d chains, in the $x$-direction. The resulting transmission probabilities are extracted at the half-filling $E=0$ when not stated otherwise. Fig.~\ref{fig:fractal_pd} displays the phase diagram in the $(M,t_2)$ plane obtained by calculation of the configuration-averaged conductance. This reveals the existence of a topological phase, which is characterized by quantized conductance $G\!=\!1$ in the thermodynamic limit, and trivial, insulating regions. As discussed below in the context of finite-size scaling, the topological phase is separated from the trivial phases by a critical line which corresponds to critical conductance $G^{c}\approx0.65$. The critical point at $(t^c_2,M^c)\!\approx\!(0,1.1)$ signifies a meeting of three distinct phases, the topological phase, a trivial spectral insulator and a trivial Anderson insulator. A finite second-nearest-neighbour hopping $|t_2|>0$ opens up a robust topological phase studied in detail below. The nature of the transition between the trivial spectral and Anderson insulator phases at the $t_2=0$ line is illustrated in Fig.~3. In Fig.~\ref{fig:fig3}(a), the mid-spectrum density of states indicates the formation of a spectral gap for $M<1.1$, while for $1.1<M<2.9$ the systems is in a gapless Anderson-localized phase. However, both phases separated by the tricritical metallic point $(t^c_2,M^c)\!\approx\!(0,1.1)$ are insulating, as seen in Fig.~\ref{fig:fig3}(b). The localization of states in the Anderson insulating phase is further analyzed by the configuration-averaged inverse participation ratio $\text{IPR}(E)=\sum_{i\alpha}|\Psi_{i\alpha}^E|^4$. The inverse participation ratio shows great enhancement around $E=0$ everywhere in the Anderson-localized regime, indicating that the spectrum is gapless but consists of trivial localized states.

\begin{figure}
\includegraphics[width=1.\columnwidth]{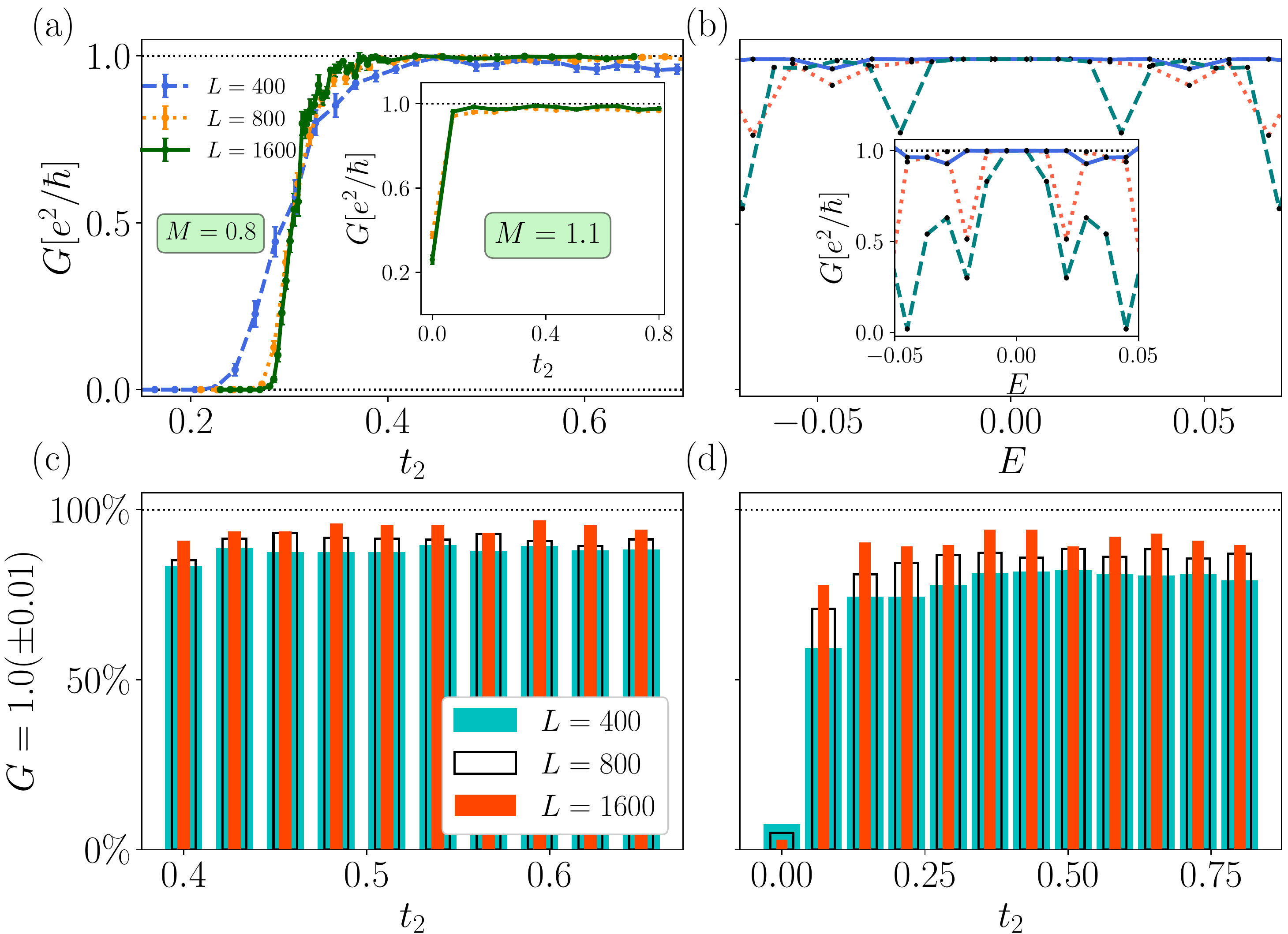}
\caption{(a) Conductance as a function of $t_2$ at $(E,M)=(0,0.8)$ signifying the transition from a spectral insulator to a topological random fractal phase. (Inset) Same for $(E,M)=(0,1.1)$. (b)  Conductance for different individual realizations of topological random fractals as a function of energy for $(t_2,M)=(0.45,0.8)$ and (inset) $(t_2,M)=(0.25,1.1)$ for $L=1600$. (c) Percentage of the samples falling into $1\%$ of the quantization range for $(E,M)=(0,0.8)$ and (d) for $(E,M)=(0,1.1)$. (c) and (d) are calculated for roughly $10^3$ independent samples at each point.}
\label{fig:fig4}
\end{figure}

\emph{Topological random fractal phase---} Having established the global features of the phase diagram, we are now establishing the properties of the topological random fractal phase. Most importantly, we concentrate on quantized conductance which is the hallmark of topological states. In contrast to the studies of deterministic fractals~\cite{fischer2021robustness,PhysRevB.101.045413,agarwala2018fractalized}, we uncover unambiguous and robust quantization of conductance in the topological random fractal phase. The formation of the quantized plateau as a function of finite $t_2$ is demonstrated in Fig.~\ref{fig:fig4}(a). At the tricritical point $(t^c_2,M^c)\!\approx\!(0,1.1)$, even a marginal increase of $t_2$ leads to formation of the topological phase with robust conductance quantization. In Fig.~\ref{fig:fig4}(b), we have plotted the conductance as a function of the energy (or the chemical potential of the leads) for a number of individual random fractal realizations. All the samples exhibit a finite quantized plateau around $E=0$.  While the topological random fractal phase is gapless, the topology is protected by a mobility gap. The width of the plateaus in Fig.~\ref{fig:fig4}(b), which show sample to sample fluctuations, corresponds to the value of the mobility gap. As long as the energy is located in the mobility gap, the conductance (for samples larger than the localization length) remains quantized.

As a testament to the remarkable robustness of the topological states, as the system size grows, the conductance quantization becomes accurate despite the great complexity and variation of different random fractal realizations. As illustrated by histograms in Fig.~\ref{fig:fig4}(c), well inside the topological regime, over 90$\%$ of configurations with the linear size $L=1600$ display conductance quantization with $1\%$ accuracy or better. As seen in Fig.~\ref{fig:fig4}(d), the quantization develops rapidly when moving from the tricritical point towards the topological phase. Larger systems exhibit on average more precise quantization, indicating that the random fractal phase is a well-defined thermodynamic phase of matter.

\begin{figure}
\includegraphics[width=1.\columnwidth]{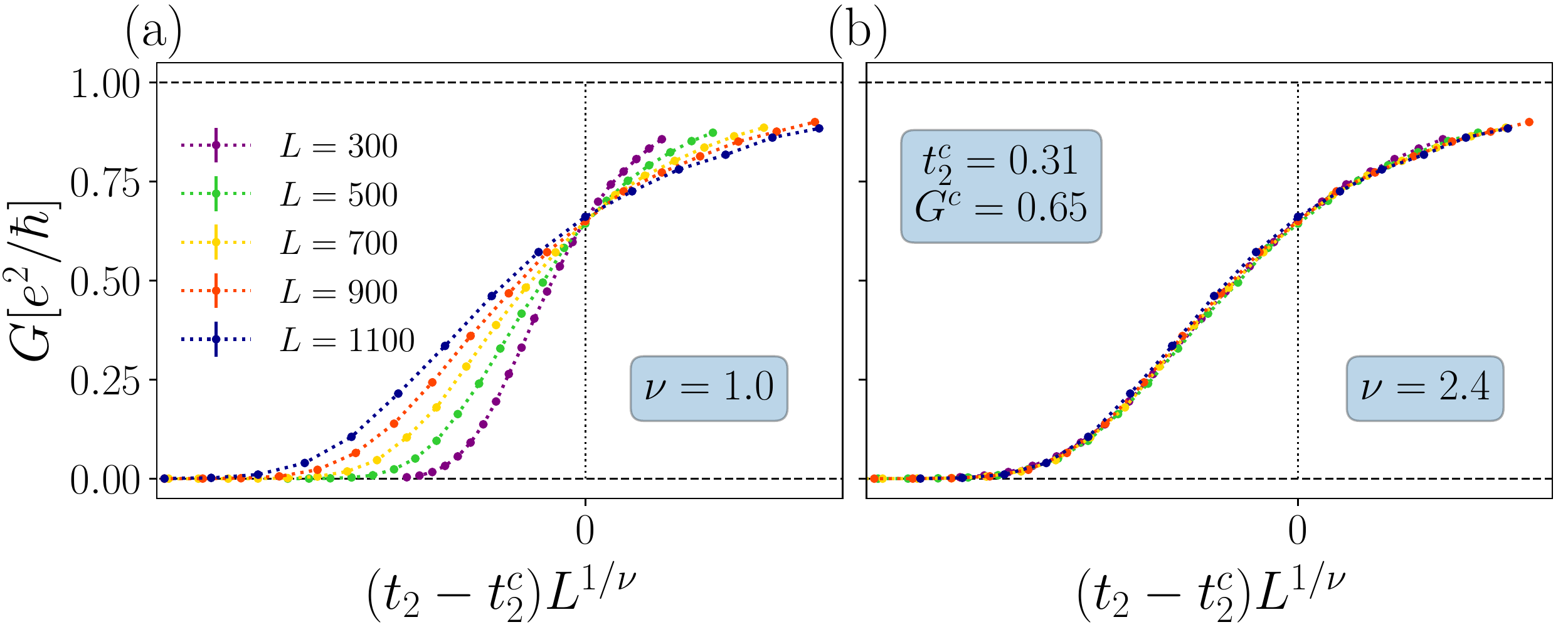}
\caption{Finite-size scaling of conductance for half-filled systems with $M=0.8$. (a) Conductance data fit with the 2d class $D$ exponent $\nu=1$ which fails to capture the correct scaling behavior. In contrast, with the exponent $\nu=2.4$ in (b) the data collapse to a universal curve. Curves are averaged over $10^3\!-\!10^4$ distinct random fractal configurations at each data point.} 
\label{fig:fig5}
\end{figure}

To illustrate that topological random fractals constitute a well-defined thermodynamic phase of matter, we carry out a finite-size scaling study. According to the theory of topological localization transitions, near the transition one expects that the configuration-averaged conductance obeys a single-parameter scaling hypothesis in the large system limit. This hypothesis predicts that the conductance curves for different system sizes collapse to a universal curve  $G=f\left[L^{1/\nu}\zeta\right]$, where $\zeta$ represents a parameter that drives the transition~\cite{huckestein1995scaling}. The scaling function $f$ approaches 0 (1) at large negative (positive) arguments. The scaling behaviour indicates that the system undergoes a sharply-defined topological phase transition at $\zeta=0$ in the thermodynamic limit, separating two distinct phases of matter. The localization length critical exponent $\nu$ is expected to be universal for all systems with the same spatial dimension and symmetry class. In particular, for symmetry class $D$ in two dimensions the exponent is $\nu = 1$~\cite{wang2021multicriticality,pan2021renormalization}. A high-precision determination of the critical exponents in the topological random fractal is beyond the scope of the present work. However, by calculating the conductance as a function of the second-nearest-neighbour hopping, to explore the validity of the scaling hypothesis $G=f\left[L^{1/\nu}(t_2-t_2^c)\right]$, we can show that the standard two-dimensional class $D$ scaling does not match the numerical evidence. In Fig.~\ref{fig:fig5}(a), we employ the value $\nu=1$ expected for the insulator-insulator phase transition for class D systems in two dimensions. The curves do accurately cross in a single point, indicating that sufficiently close to the critical point, the system sizes $L=300-1100$ are in the single-parameter scaling regime, but the data clearly do not follow a single curve. As a contrast, as seen in Fig.~\ref{fig:fig5}(b) for a higher value exponent $\nu=2.4$, the curves collapse to a single curve near the critical point. While not yielding a high-precision numerical value for the exponent, the data supports the conclusion that the transition obeys scaling behaviour. The observed substantial departure of the critical exponent from its universal 2d value further suggests that, despite bearing similarities to its integer-dimensional counterpart, the random fractal phase is a genuinely distinct phase of matter with unique critical properties. Additionally, as the curves are accurate near the critical point we can extract the critical conductance  $G^c\approx 0.65$, which provides the basis for the white dashed line as an approximate phase boundary in Fig.~\ref{fig:fractal_pd}.

\emph{Summary and outlook---} In this work we introduced a new electronic state of matter, topological random fractals, and established its central properties. Being the most complex realization of nontrivial band topology known to date, they support robust quantized conductance protected by a mobility gap. The finite-size scaling results suggest that the topological random fractals belong to a different universality class than their integer-dimensional parent states, calling for further studies on topological fractals. 
Besides the fundamental interest, there is reason for optimism that such systems will become available for experimental studies in the near future. Technological advances have enabled fabrication of artificial and quantum simulator systems realizing quasicrystalline and fractal electronic structures. These advances suggest that experimental realization of topological random fractals may not be far behind.

\emph{Acknowledgements-- } The authors acknowledge the Academy of Finland project 331094 for support.

\bibliography{topofractal}

\end{document}